\documentclass{Interspeech2024}

\interspeechcameraready 

\usepackage{multirow}
\usepackage{diagbox}
\usepackage{tabularx}
\title{EmoSphere-TTS: Emotional Style and Intensity Modeling via Spherical Emotion Vector for Controllable Emotional Text-to-Speech}

\name[affiliation={1}]{Deok-Hyeon}{Cho}
\name[affiliation={1}]{Hyung-Seok}{Oh}
\name[affiliation={1}]{Seung-Bin}{Kim}
\name[affiliation={2}]{Sang-Hoon}{Lee}
\name[affiliation={{\dagger}1}]{Seong-Whan}{Lee}

\address{
  $^1$Department of Artificial Intelligence, Korea University, Seoul, Korea \\
  $^2$Department of Software and Computer Engineering, Ajou University, Gyeonggi-do, Korea \
  \thanks{$^\dagger$Corresponding author}
\email{dh\_cho@korea.ac.kr, hs\_oh@korea.ac.kr, sb-kim@korea.ac.kr, sanghoonlee@ajou.ac.kr, sw.lee@korea.ac.kr}}

\keywords{Text-to-speech, expressive emotional speech synthesis, emotional style and intensity control}

\begin{document}
\maketitle
\begin{abstract}
Despite rapid advances in the field of emotional text-to-speech (TTS), recent studies primarily focus on mimicking the average style of a particular emotion. As a result, the ability to manipulate speech emotion remains constrained to several predefined labels, compromising the ability to reflect the nuanced variations of emotion. In this paper, we propose EmoSphere-TTS, which synthesizes expressive emotional speech by using a spherical emotion vector to control the emotional style and intensity of the synthetic speech. Without any human annotation, we use the arousal, valence, and dominance pseudo-labels to model the complex nature of emotion via a Cartesian-spherical transformation. Furthermore, we propose a dual conditional adversarial network to improve the quality of generated speech by reflecting the multi-aspect characteristics. The experimental results demonstrate the model’s ability to control emotional style and intensity with high-quality expressive speech.
\end{abstract}

\section{Introduction}
Recently, text-to-speech (TTS) has shown rapid progress \cite{wang17n_interspeech, lee2022hierspeech, 10381805}, and interest in emotional TTS has also increased. Although emotional TTS technology has seen significant improvements in recent years, there remains a limitation in high-level interpretable emotion control \cite{wang23ka_interspeech, liu23t_interspeech, liu23u_interspeech, kang23_interspeech}. In particular, emotional control remains challenging, as speech labeled with the same emotion can exhibit diverse emotional expressions greatly influenced by the variability in acting performances.

For emotional TTS, a common approach is to control the diverse emotional expressions from the emotion labels and the reference audio. The emotion label-based approach models the complex nature of emotion to serve as an auxiliary input for the TTS system. Relative attribute \cite{zhou2023speech, zhu2019controlling, zhou2022emotion} is one of the most representative methods, utilizing a learned ranking function \cite{parikh2011relative} to delineate differences between binary classes. EmoQ-TTS \cite{im2022emoq} employs distance-based quantization via linear discriminant analysis to control fine-grained emotional intensity. Another way to control the diverse expression of emotion is through reference-based emotional TTS. In this context, a scaling factor \cite{li22h_interspeech, li2022cross} is multiplied by the emotion embedding to control emotion intensity precisely. However, these methods have several limitations. All methods based on emotion labels focus on transferring emotions using discrete labels that ignore the complex nature of emotion conveyed in human speech \cite{shin22b_interspeech}. For example, while $sad$ spans emotions like $lonely$ and $hurt$, its categorization typically reduces expressions to a uniform style. Using references to control emotional expression is further complicated by the difficulty in capturing the nuances of references due to a mismatch between the reference and the synthesized speech. Furthermore, these methods face difficulty finding suitable scaling factors, and making adjustments often results in unstable audio quality.

Another strategy to control emotional expression involves leveraging the emotional dimensions. Russell \cite{russell1980circumplex} proposed a continuous emotion space as an alternative for human emotions. Building on this, researchers attempt to control emotional expression by utilizing the extended emotional dimensions of arousal, valence, and dominance (AVD) \cite{habib2019semi, sivaprasad21_interspeech}. Emotional dimensions provide a continuous and fine-grained description, offering more detailed control than discrete emotions. However, only a few emotional speech databases provide these annotations due to the inherent subjectivity and the high costs associated with collecting such data. Furthermore, emotional dimensions are challenging for models to control intuitively and distinctly characterize the diverse emotional expression.

To address the above limitations, we propose EmoSphere-TTS, a novel approach to control emotional style and intensity with spherical emotion vector space in emotional TTS. We adopt the emotional dimensions of AVD from pseudo-labeling in speech emotion recognition (SER). We also propose a spherical emotion vector space via Cartesian-spherical transformation to model the complex nature of emotion that has been difficult in Cartesian coordinate systems. We found that this space is the key to controlling the emotional style and intensity of the synthetic speech. Furthermore, we introduce dual conditional adversarial training to improve the quality of generated speech by reflecting the emotion and speaker-specific characteristics. The experimental results demonstrate that our model outperforms the others in terms of audio quality and emotion similarity on the controllable emotional TTS. Audio samples are available at \url{https://EmoSphere-TTS.github.io/}.

\section{EmoSphere-TTS}
We present a controllable emotional TTS system, EmoSphere-TTS. We introduce spherical emotion vector space and spherical emotion encoder to deliver speech with the complex nature of emotion. Furthermore, we introduce a dual conditional discriminator for better audio quality and expressiveness. The details are described in the following subsections.

\begin{figure*}[!t] 
    \centering
\includegraphics[width=0.98\linewidth]{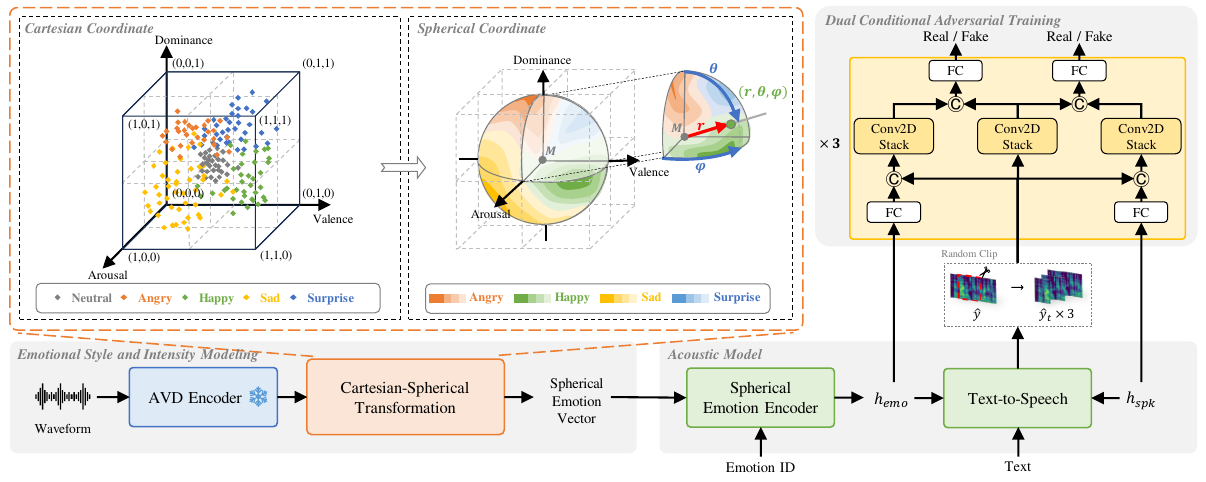}\vspace{-0.3cm}
\caption{Overall architecture of EmoSphere-TTS.}
\label{model}\vspace{-0.3cm}
\end{figure*}

\subsection{Emotional style and intensity modeling}
In this section, we model the diverse emotional expressions through a spherical emotion vector space. The approach to building the space is structured around two key components: i) the AVD encoder and ii) the Cartesian-spherical transformation.

\subsubsection{AVD encoder}
Instead of using emotional dimensions of human annotation, we adopt wav2vec 2.0 \cite{baevski2020wav2vec}-based SER \cite{wagner2023dawn} to extract consistently continuous and detailed representations from audio. The model generates predictions for $e_{ki}=(d_{a},d_{v},d_{d})$, where $d_{a}$ represents arousal, $d_{v}$ valence, and $d_{d}$ dominance, each ranging approximately from 0 to 1 in Cartesian coordinates. Here, $e_{ki}$ denotes the $i$-th coordinate of the $k$-th emotion.

\subsubsection{Cartesian-spherical transformation}
For modeling the complex nature of emotion, we introduce the spherical emotion vector space, which represents the relative distance and angle vector from the neutral center. Inspired by coordinate transformations \cite{jenke2018cognitive}, which can be easily controlled by a continuous scalar indicating emotional style and intensity. We transform all points for AVD pseudo-labels in spherical coordinates by following the assumptions: i) the emotional intensity increases as it moves farther from the neutral emotion center, and ii) the angle from the neutral emotion center determines the emotional style. First, we obtain transformed Cartesian coordinates $e'_{ki}=(d'_{a},d'_{v},d'_{d})$ by setting the neutral emotion center $M$ as the origin,
\begin{equation}
e'_{ki} = e_{ki} - M \, \text{   where   } \, M = \frac{1}{N_{n}} \sum_{i=1}^{N_{n}} e_{ni},
\end{equation}
where $N_{n}$ represents the total number of neutral coordinates $e_{ni}$. Then, the transformation from Cartesian coordinates to spherical coordinate $(r,\vartheta,\varphi)$ can be formulated as:
\begin{equation}
r = \sqrt{{d'_{a}}^{2} + {d'_{v}}^{2} + {d'_{d}}^{2}},
\end{equation}
\begin{equation}
\vartheta = \arccos\left(\frac{d'_{d}}{r}\right), \varphi = \arctan\left(\frac{d'_{v}}{d'_{a}}\right).
\end{equation}
After the Cartesian-spherical transformation, we normalize the intensity of the emotion by scaling the radial distance $r$ to a range of $0$ to $1$. To achieve this, the min-max normalization process utilizes the interquartile range technique \cite{walfish2006review}, effectively determining the minimum and maximum values for the scale. Additionally, we quantize the emotion style by segmenting directional angles $\vartheta$ and $\varphi$ into eight octants, each defined by the positive and negative directions of the $A$, $V$, and $D$ axes.

\begin{table*}[!ht]
    \centering
        \caption{Subjective and objective evaluation results. The last group of rows represents an ablation study of the proposed model. The nMOS and sMOS scores are presented with 95\% confidence intervals.}
    \label{Table1}\vspace{-0.3cm}
        \resizebox{0.98\textwidth}{!}{
    \begin{tabular}{l|c|cc|cc|ccc|cc|c}
        \toprule
        \textbf{Method} & \textbf{UTMOS} ($\uparrow$) & \textbf{nMOS} ($\uparrow$) & \textbf{sMOS} ($\uparrow$) & \textbf{WER} ($\downarrow$) & \textbf{CER} ($\downarrow$) & $\textbf{RMSE}_{f0}$ ($\downarrow$) & $\textbf{RMSE}_{period}$ ($\downarrow$) & \textbf{F1 {V/UV}} ($\uparrow$) & \textbf{EER} ($\downarrow$) & \textbf{SECS} ($\uparrow$) & \textbf{ECA} ($\uparrow$) \\ 
        \midrule
            GT & 3.78 & 4.27$\pm$0.04 & 4.37$\pm$0.11 & 11.92 & 3.04 & - & - & - & 2.99 & 0.753 & 85.67 \\ 
            BigVGAN \cite{lee2023bigvgan} & 3.62 & 4.26$\pm$0.04 & 4.30$\pm$0.11 & 11.97 & 3.03 & 32.21 & 0.214 & 0.8630 & 3.19 & 0.741 & 86.07   \\ 
        \midrule
            FastSpeech 2 w/ Emotion Label \cite{ren2020fastspeech} & 2.52 & 3.13$\pm$0.06 & 3.28$\pm$0.15 & 21.09 & 9.30 & 44.65 & 0.455 & 0.6995 & 5.89 & 0.654 & 93.75 \\ 
            FastSpeech 2 w/ Relative Attribute \cite{zhu2019controlling} & 2.57 & 3.16$\pm$0.06 & 3.22$\pm$0.14 & 23.70 & 10.02 & 44.82 & 0.454 & 0.6978 & 6.20 & 0.654 & 92.80 \\ 
            FastSpeech 2 w/ Scaling Factor \cite{li2022cross} & 2.66 & 3.34$\pm$0.05 & 3.29$\pm$0.13 & 19.46 & 8.73 & \textbf{41.90} & 0.453 & 0.7028 & 6.39 & 0.645 & 64.19  \\ 
        \midrule
            EmoSphere-TTS (Proposed) & \textbf{3.15} & \textbf{3.88}$\pm$\textbf{0.05} & \textbf{3.48}$\pm$\textbf{0.11} & \textbf{17.43} & \textbf{7.05} & 42.57 & \textbf{0.447} & 0.7077 & \textbf{4.29} & \textbf{0.669} & \textbf{94.02} \\ 
            w/o Spherical Emotion Vector & 3.04 & 3.69$\pm$0.05 & 3.45$\pm$0.12 & 19.22 & 8.05 & 43.70 & 0.449 & \textbf{0.7099} & 5.50 & 0.649 & 93.89 \\ 
            w/o Dual Conditional Discriminator & 2.86 & 3.39$\pm$0.06 & 3.24$\pm$0.14 & 18.09 & 7.38 & 42.52 & 0.449 & 0.7055 & 5.86 & 0.666 & 92.60 \\ 
        \bottomrule
    \end{tabular}
      }\vspace{-0.4cm}
\end{table*}

\subsection{Spherical emotion encoder}
After building spherical emotion vector space, the spherical emotion encoder blends them with emotion ID to compose their spherical emotion embedding. Initially, we use a projection layer to align the dimensions of the emotion style vector and emotion class embedding. Then, we concatenate these projections and apply a softplus activation \cite{zheng2015improving} similar to \cite{yoon22b_interspeech} followed by Layer Normalization (LN) \cite{ba2016layer}. Finally, the spherical emotion embedding, $\mathbf{h}_{emo}$, is merged with projected emotion intensity vectors, as the following equation:
\begin{equation}
\mathbf{h}_{emo} = \text{LN} \left( \text{softplus}\left( \text{concat} \left(\mathbf{h}_{sty}, \mathbf{h}_{cls}\right) \right) \right) + \mathbf{h}_{int}.
\end{equation}
Here, $\mathbf{h}_{sty}$, $\mathbf{h}_{int}$, and $\mathbf{h}_{cls}$ denote the outputs of the projection layer related to the emotional style vector, emotional intensity vector, and emotion class embedding, respectively.

\subsection{Dual conditional adversarial training}
We adopt the structure of multiple CNN-based discriminators \cite{ye2022syntaspeech} for adversarial training to improve the quality of the TTS model. These discriminators comprise a Conv2D stack that consists of multiple stacked 2D-convolutional layers and fully connected (FC) layers. The input value is the random Mel-spectrogram clip (Mel clip) using random windows of different lengths $t$. To improve quality and further expressiveness, we utilize emotion and speaker embeddings to capture the multi-aspect characteristics more effectively, inspired by \cite{yang21e_interspeech, 10517426}. One Conv2D stack receives only the Mel clip, while the others receive a combination of condition embedding and the Mel clip. We extend the condition embedding to match the length of the Mel clip for concatenation. The loss functions $\mathcal{L}$ of the discriminator $D$ and generator $G$ are shown in Equation (\ref{D_loss}) and (\ref{G_loss}):
\begin{equation}
    \mathcal{L}_{D}=\sum_{c \in \{spk, emo\}}\sum_{t}\mathbb{E}[(1-D_{t}(y_t,c))^2+ D_{t}(\hat{y}_t,c)^2], 
    \label{D_loss}
\end{equation}
\begin{equation}
\mathcal{L}_{G}=\sum_{c \in \{spk, emo\}}\sum_{t}\mathbb{E}[(1-D_{t}(\hat{y}_t,c))^2],
\label{G_loss}
\end{equation}
where $y_{t}$ and $\hat{y}_t$ respectively represent the ground truth and generated Mel-spectrograms, with $c$ denoting the condition type.

\subsection{TTS model}
We retain the original architecture and objective function of FastSpeech 2 \cite{ren2020fastspeech} except for using an emotion spherical vector to provide emotional style and intensity information. Additionally, the speaker ID is mapped into an embedding $h_{spk}$ to represent different speaker characteristics. Then, the speaker and emotion embedding are concatenated and provided to the variance adaptor. During inference, we use manual style and intensity vectors to control the diverse emotional expressions. By manipulating the emotional style and intensity in the spherical emotion vector space, we can efficiently synthesize the complex nature of emotion and control the diverse emotional expressions in synthesized speech.

\section{Experiments and results}

\subsection{Experimental setup}
We use the emotional speech dataset (ESD) \cite{zhou2022emotional}, which consists of about 350 parallel utterances spoken by 10 English speakers with five emotional states (neutral, happy, angry, sad, and surprise). Following the prescribed data partitioning criteria, we extracted one sample for each emotion from every speaker, resulting in a total of 17,500 samples. For the Mel-spectrogram, we transform audio using the short-time Fourier transform with a hop size of 256, a window size of 1,024, an FFT size of 1,024, and 80 bins of Mel-filter. We employ the AdamW optimizer \cite{loshchilov2017decoupled}, setting the hyperparameters $\beta_{1}$ to 0.9 and $\beta_{2}$ to 0.98. For the training of the TTS system and discriminator, the learning rates were configured at $5\times 10^{-4}$ and $1\times 10^{-4}$, respectively. The training process of the TTS module was conducted over approximately 24 hours on a single NVIDIA RTX A6000 GPU. For the audio synthesis in our experiments, we utilize the official implementation of BigVGAN \cite{lee2023bigvgan}, along with its pre-trained model.

\subsection{Implementation details}
For the acoustic model, following the FastSpeech 2 \cite{ren2020fastspeech} configuration, in the FFT block of the phoneme encoder and decoder, we configure the number of layers to 4, hidden size to 256, filter size to 1024, and kernel size to 9. Regarding the AVD encoder, we adopt a system proposed in \cite{wagner2023dawn}, which predicts AVD using wav2vec 2.0 \cite{baevski2020wav2vec} and a linear predictor. For the discriminator, we use projection layers for each condition with a hidden size of 128 and three different sizes of windows ([32, 64, 96]).

\begin{figure*}[!t] 
    \centering
\includegraphics[width=1\linewidth]{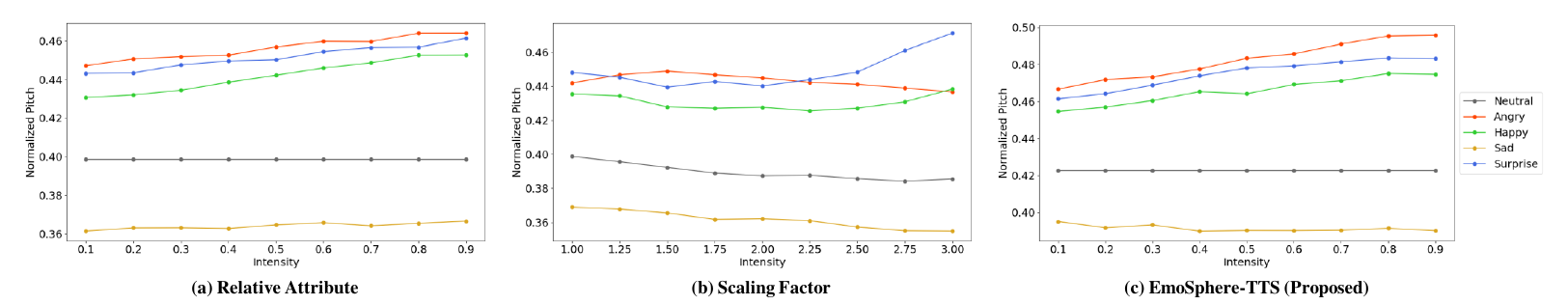}\vspace{-0.25cm}
\caption{The pitch tendency according to emotion and intensity.}
\label{Intensity_control}\vspace{-0.3cm}
\end{figure*}

\begin{figure*}[!t] 
    \centering
\includegraphics[width=1\linewidth]{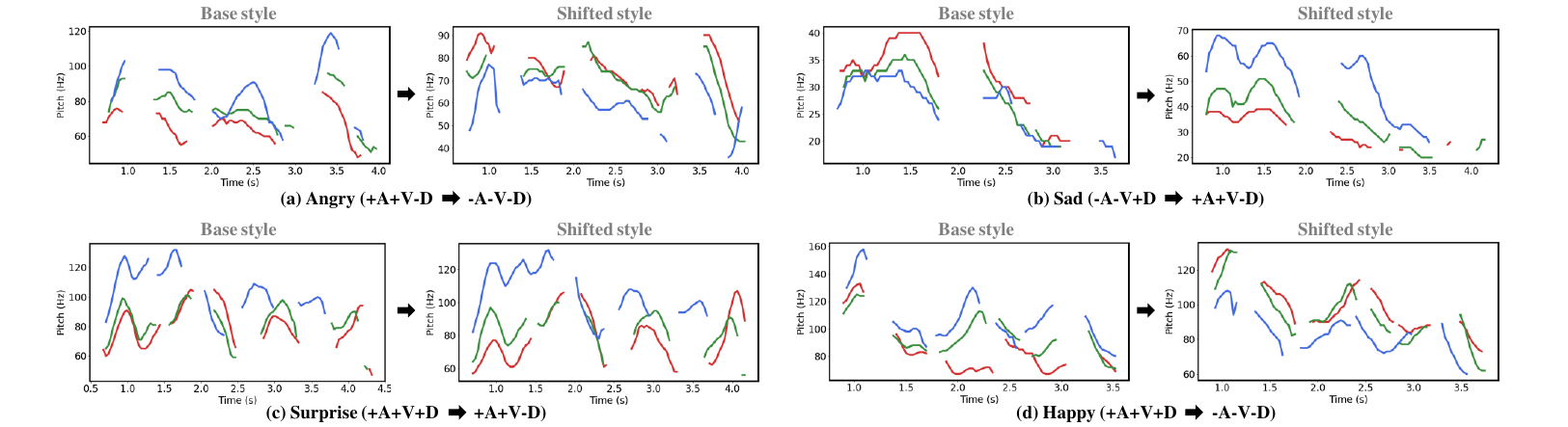}\vspace{-0.25cm}
\caption{A pitch track of a sample demonstrating the effects of emotional style shift, where $A$, $V$, and $D$ represent arousal, valence, and dominance, respectively. The line color represents emotional intensity, red = 0.1, green = 0.5, and blue = 0.9.}
\label{Style_control}\vspace{-0.3cm}
\end{figure*}

\subsection{Model performance}
We compare EmoSphere-TTS with other systems: 1) FastSpeech 2 w/ emotion label \cite{ren2020fastspeech}, which adopts emotion id with a look-up table. 2) FastSpeech 2 w/ relative attribute \cite{zhu2019controlling}, which assigns emotion intensity via learned ranking function \cite{parikh2011relative}. 3) FastSpeech 2 w/ scaling factor \cite{li2022cross}, which is multiplied by the emotion embedding to control emotion intensity precisely. We follow the same experimental setup and model configurations to ensure a fair comparison. For evaluation metrics, we use a naturalness mean opinion score (nMOS) and a similarity mean opinion score (sMOS). The nMOS and sMOS are reported with 95\% confidence intervals. Additionally, we utilize the open-source UTMOS\footnote{https://github.com/tarepan/SpeechMOS} \cite{saeki22c_interspeech} as a MOS prediction model for the naturalness metric. To evaluate linguistic consistency, we calculate the word error rate (WER) and character error rate (CER) by Whisper large model \cite{radford2023robust}. For the speaker similarity measurements, we calculate the speaker embedding cosine similarity (SECS) via Resemblyzer\footnote{https://github.com/resemble-ai/Resemblyzer} between the target and converted speech and equal error rate (EER) via the automatic speech recognition model \cite{kwon2021ins}. For emotionally expressive evaluation, we conduct emotion classification accuracy (ECA) using a pre-built emotion classification model \cite{9778970}. For prosodic evaluation, we compute the root mean square error for both pitch error (RMSE$_{f_0}$) and periodicity error (RMSE$_{period}$), along with the F1 score of voiced/unvoiced classification (F1$_{v/uv}$).

\begin{table}[!t]
    \caption{Results of evaluation for emotion intensity control. Emotion Intensity Recognition is asked to select the sample with the stronger intensity from a pair. Weak to Strong represents emotional intensity.}
    \label{Table2}\vspace{-0.3cm} 
    \centering
        \resizebox{1\columnwidth}{!}{
    \begin{tabular}{c|c|c|c|c}
    \toprule
        \multirow{2}*{\textbf{Emotion}} & \multirow{2}*{\textbf{Method}} & \multicolumn{3}{c}{\textbf{Emotion Intensity Recognition [\%]}} \\ 
        \cline{3-5} & & \textbf{Weak$<$Medium} & \textbf{Medium$<$Strong} & \textbf{Weak$<$Strong} \\
    \midrule
        \multirow{3}*{Angry}    & Relative Attribute \cite{zhu2019controlling} & 0.56 & 0.56 & 0.72 \\ 
                                & Scaling Factor \cite{li2022cross} & 0.44 & 0.52 & 0.48 \\ 
                                & \textbf{EmoSphere-TTS} & \textbf{0.72} & \textbf{0.75} & \textbf{0.79} \\ 
    \midrule
        \multirow{3}*{Sad}      & Relative Attribute \cite{zhu2019controlling} & 0.57 & 0.39 & 0.55 \\ 
                                & Scaling Factor \cite{li2022cross} & \textbf{0.69} & \textbf{0.48} & \textbf{0.63} \\ 
                                & \textbf{EmoSphere-TTS} & 0.62 & 0.42 & 0.48 \\ 
    \midrule
        \multirow{3}*{Happy}    & Relative Attribute \cite{zhu2019controlling} & 0.48 & 0.51 & 0.60 \\ 
                                & Scaling Factor \cite{li2022cross} & 0.61 & 0.43 & 0.44 \\ 
                                & \textbf{EmoSphere-TTS} & \textbf{0.80} & \textbf{0.66} & \textbf{0.84} \\ 
    \midrule
        \multirow{3}*{Surprise} & Relative Attribute \cite{zhu2019controlling} & 0.44 & 0.55 & 0.52 \\ 
                                & Scaling Factor \cite{li2022cross} & 0.50 & 0.45 & 0.46 \\ 
                                & \textbf{EmoSphere-TTS} & \textbf{0.69} & \textbf{0.76} & \textbf{0.79} \\ 
    \midrule
    \midrule
        \multirow{3}*{Average}  & Relative Attribute \cite{zhu2019controlling} & 0.52 & 0.51 & 0.60 \\ 
                                & Scaling Factor \cite{li2022cross} & 0.56 & 0.47 & 0.50 \\ 
                                & \textbf{EmoSphere-TTS} & \textbf{0.71} & \textbf{0.65} & \textbf{0.72} \\ 
    \bottomrule
    \end{tabular}
    }\vspace{-0.3cm}
\end{table}

Given our primary on quality and expressiveness, manual intensity and style vectors are not employed in these experiments. In the ablation study, w/o spherical emotion vector is a model that uses emotion id with a lookup table instead of a spherical emotion vector and encoder. As shown in Table \ref{Table1}, our model achieves significant improvements, and this can be explained by: 1) as opposed to transferring emotion from emotion label-based or reference audio methods, directly assigning spherical emotion vector is easier for the model to generate good quality speech. Our spherical emotion vector exhibits better expressiveness and audio quality, including naturalness and pronunciation, even without a dual conditional discriminator. 2) The dual conditional discriminator improves the quality of the generated speech by reflecting both emotion and speaker characteristics.

\subsection{Emotion intensity controllability}
In this section, we conduct a subjective evaluation to determine the discernibility of synthesized speech samples exhibiting varying intensity levels. To demonstrate the intensity control capability of our model, we synthesize speech with three different levels of emotion intensity (weak, medium, and strong). Evaluators are presented with two different sentences, each with varying intensities, and tasked with selecting the one that exhibits stronger emotion. In relative attribute \cite{zhu2019controlling} and EmoSphere-TTS, we uniformly refer to scores 0.1 as weak, 0.5 as medium, and 0.9 as strong. Scaling factor \cite{li2022cross} cannot assign intensity scores, so we set the scalar factor to 1, 2, and 3 to represent the weak, medium, and strong emotion strength the same as in the original setting. Additionally, we visualize the tendency of pitch to demonstrate the ability to control emotional expression as shown in Figure \ref{Intensity_control}. We computed these values by averaging the pitch from the synthesized speeches by all combinations of emotion labels and intensity vectors for all test sentences. 

As shown in Table \ref{Table2}, relative attribute \cite{zhu2019controlling} effectively controls the intensity. However, in the $sad$ emotional speech, the pitch increases as the intensity increases, as shown in Figure \ref{Intensity_control} (a). This indicates that subtle emotional nuances are complex to capture when considering emotion labels alone, and expressions are often reduced to a uniform style. Scaling factor \cite{li2022cross} might not be efficient for performing intensity control; in some cases, the intensity difference is not readily perceivable, as shown in Figure \ref{Intensity_control} (b). However, as shown in Table 2, the scaling factor outperformed the other models for the $sad$ emotion. Still, the scaling factor focuses on reducing pitch and decelerating speech rate in static emotions, overlooking the complex nature of emotion. On the other hand, compared to the baseline models, EmoSphere-TTS performs the best. Furthermore, the pitch tendency plot reflects the intensity according to the emotion. This indicates that the proposed model synthesizes speech according to a given intensity scale.

\subsection{Emotion style shift}
To demonstrate the changing patterns of emotion intensity based on the shifted emotion style, we visualize the pitch tracks of a sample. In Figure \ref{Style_control}, we observe that when the base style vector is input, emotion intensity changing patterns reflect the characteristics of the AVD axes. For example, style vectors with positive $A$ axes have a pitch that tends to increase the changing patterns, positive $V$ axes have higher average pitch values, and positive $D$ axes have a narrow range in changing patterns. This indicates that arousal, valence, and dominance reflect each meaning of axes, representing the level of excitement or energy, positivity or negativity of emotion, and control level within an emotional state, respectively. By shifting the style vector, the emotion intensity patterns change with the shifted axis. This indicates that the proposed spherical emotion vector reflects diverse emotional expressions and offers detailed manipulation of emotional expressions.

\section{Conclusion}
We present EmoSphere-TTS, a system that synthesizes expressive emotional speech through a spherical emotion vector space, controlling the diverse emotion expression. With only a speech dataset, we extract AVD pseudo-labels and model generalized representations of the emotional style and intensity through Cartesian-spherical transformation. Furthermore, we improve the quality and emotional expressiveness of the overall model using the dual conditional adversarial discriminator and spherical emotion encoder. The experimental results demonstrate that our proposed spherical emotion vector effectively synthesizes the complex nature of emotion and controls diverse emotion expression. In this article, we only focused exclusively on the global style present within sentence-level emotional information. In future work, we aim to extend our approach to include phoneme-level emotional information and allow for fine-grained control. We also expect that the proposed method can be utilized for emotional voice conversion like \cite{9778970, lee23i_interspeech, oh2024durflex}. 

\section{Acknowledgements}
This work was partly supported by Institute of Information \& Communications Technology Planning \& Evaluation (IITP) grant funded by the Korea government (MSIT) (No. 2019-0-00079, Artificial Intelligence Graduate School Program (Korea University), No. 2021-0-02068, Artificial Intelligence Innovation Hub, and AI Technology for Interactive Communication of Language Impaired Individuals).

\bibliographystyle{IEEEtran}
\bibliography{refs}

\end{document}